\documentclass[twocolumn,preprintnumbers,amsmath,amssymb,a4paper,10pt]{revtex4}
\usepackage{amsthm}
\usepackage{mathrsfs}
\usepackage{graphicx}
\usepackage{dcolumn}
\usepackage{bm}
\usepackage{amsmath}
\usepackage{subfigure}

\usepackage[permil]{overpic}
\begin{document}

\title{Experimental demonstrations of high-Q superconducting coplanar waveguide resonators}

\author{LI HaiJie$^{1}$, WANG YiWen$^{1}$, WEI LianFu$^{1,2}$\footnote{E-mail: weilianfu@gmail.com}, Zhou PinJia$^{1}$, WEI Qiang$^{1}$, CAO ChunHai$^{3}$, FANG YuRong$^{3}$, YU Yang$^{3}$, WU PeiHeng$^{3}$}

\address{1. Quantum
Optoelectronics Laboratory, 
Southwest Jiaotong University, Chengdu 610031, China \\ 2. State Key
Laboratory of Optoelectronic Materials and Technologies, School of
Physics and Engineering, Sun Yat-Sen University, Guangzhou 510275,
China\\ 3. Research Institute of Superconductor Electronics, Nanjing
University, Nanjing 210093, China}

\begin{abstract}
We designed and successfully fabricated an absorption-type of
superconducting coplanar waveguide (CPW) resonators. The resonators
are made from a Niobium film (about 160 nm thick) on a
high-resistance Si substrate, and each resonator is fabricated as a
meandered quarter-wavelength transmission line (one end shorts to
the ground and another end is capacitively coupled to a through
feedline). With a vector network analyzer we measured the
transmissions of the applied microwave through the resonators at
ultra-low temperature (e.g., at $20$ mK), and found that their
loaded quality factors are significantly high, i.e., up to $\sim
10^{6}$. With the temperature increases slowly from the base
temperature (i.e., $20$ mK), we observed the resonance frequencies
of the resonators are blue shifted and the quality factors are
lowered slightly. In principle, this type of CPW-device can
integrate a series of resonators with a common feedline, making it a
promising candidate of either the data bus for coupling the distant
solid-state qubits or the sensitive detector of single photons.
\\
\\
\textbf{\emph{keywords}}: \\Superconducting coplanar waveguide
resonator, resonance frequency, quality factor
\end{abstract}

\maketitle

Superconducting coplanar waveguide (CPW) resonators have been
extensively studied for years, specifically as various sensors
including radiation detectors~\cite{FSTC1,FSTC2,FSTC3}, parameter
amplifiers~\cite{CLFD1}, and the data bus of superconducting
qubits~\cite{LZBT1,LZBT2,LZBT3}, etc.. Compared to other low
temperature detectors, e.g., transition edge sensors (TESs)
\cite{TES1,TES2}, superconducting tunnel junctions (STJs)
\cite{STJ3}, one of the advantages of the CPW device is its
integrability and relatively-easy frequency domain multiplexing. In
principle, a series of CPW resonators with different frequencies can
be measured by using a single feedline. This greatly simplifies the
system structure and the relevant cryogenic electronics.

Particularly, a superconducting CPW resonator with sufficiently-high
quality factor can service as a sensitive radiation detector.
Physically, the incoming photons break Cooper pairs in the
superconducting resonator and create quasiparticle excitations. This
changes the kinetic inductance~\cite{KI1} and resonance frequency of
the CPW resonator, which can be easily measured experimentally. In
principle, higher quality factor allows more sensitive detection of
photons.
Also, CPW resonators with high quality factors (i.e., with narrower
3dB-bandwidths) imply that more resonators (with different resonant
frequencies) can be multiplexed on-chip for a fixed bandwidth. This
is essentially important for the development of large sensor-arrays
used in astronomy~\cite{FSTC1,FSTC6}.

Due to the potentially-excellent sensitivity and various attractive
applications, absorbed-type superconducting resonators resonators
have been paid much attention in recent
years~\cite{FSTC1,FSTC2,FSTC3,FSTC6}. Here, we report an
experimental demonstration of the desirable meandered
quarter-wavelength transmission line resonators, which can be used
as the absorbed-type Microwave Kinetic Inductance Detectors
(MKIDs)~\cite{FSTC1,FSTC6} with high quality factors.
Our device is made from a Nb film ($t\sim 160$nm) deposited on a
high-resistance Si subtracts by magnetron sputtering. The designed
circuit is generated by the usual photolithography. The measured
loaded quality factor of our resonator at 20 mK is $Q_{l}$ = 1.1654
$\times 10^{6}$ for the resonant frequency $f_0=1.8575$GHz. We
alsomeasured the temperature dependence of the resonance frequency
in the range of base temperature to around 1180 mK.

\section{Device and measurement system}
The microscope photograph of our meandered CPW $\lambda$/4
transmission line resonator is shown in Fig.~1. The resonator shorts
to the ground plane at one end,  and capacitively couples to a
through line at another end.
In our design, niobium is selected as the metal film material, due
to its well superconductivity at low temperature. The characteristic
impedance $Z_0$ of through line is designed to be $50$ $\Omega $, in
order to avoid the reflections of the driving signals. In fact, for
a single layer dielectric the impedance can be analytically
calculated by using the formula~\cite{Brian C. Wadell}
\begin{equation}
Z_{0}=\frac{30\pi}{\sqrt{\epsilon _{e,t}}}\frac{K(k')}{K(k)}.
\end{equation}
Here, $\epsilon _{e,t}$ is the effective dielectric constant with a
correction of the film thickness t,
$K(k)=\int_0^{\pi/2}d\theta/\sqrt{1-k^2\sin^2\theta}$ the complete
elliptic integral with modulus $k$. The parameters $k$ and $k'$ are
determined by the resonator's geometry parameters~\cite{Brian C.
Wadell}: the center strip width $w$, slot width between the center
strip and the ground plane $s$, metal film thickness $t$ and the
thickness $h$ of the Silicon substrate.
%
Physically, the fundamental resonance frequency $f_{0}$
is determined by the overall length of the CPW resonator $l$ and the
effective substrate dielectric parameter $\epsilon$.
\begin{equation}
f_{0}=\sqrt{\frac{2}{\epsilon+1}}\frac{c}{4l},
\end{equation}
Above, $c$ is the speed of an electromagnetic wave in a vacuum.
In our design, the length of the CPW resonator is set as $15.814$
mm, corresponding to a fundamental resonance frequency $f_0=1.8575$
GHz. Also, the center strip width $w_{t}$ is set as $20$\,$\mu$m and
the slot width $s_{t}$ between the center strip and the ground plane
is $15$\,$\mu$m. Note that the CPW resonator is decoupled from the
through line so its impedance is not necessary to be 50 $\Omega$.

Several aspects, e.g., coupling strength between the through line
and the resonator, film and dielectric losses, and also the
radiation loss, etc., may influence the quality factor $Q$ of the
CPW resonator.
For the present device, i.e., niobium film deposited on the Silicon
substrate, the film and dielectric losses are very small at very low
temperature ($T\ll T_{c}, T_{c}=9.2$K for niobium). Also, radiation
loss is negligible, as the line width of the transmission line is
very narrow. As a consequence, at $T\ll T_{c}$ the quality factor
$Q_{c}$, related mainly to the coupling capacitance, dominates the
loaded (total) quality factor $Q_{l}$ defined by
\begin{equation}
\frac{1}{Q_{l}}=\frac{1}{Q_{i}}+\frac{1}{Q_{c}}.
\end{equation}
Here, $Q_{i}$ is the internal (unloaded) quality factor, accounting
for the energy leakage due to all other losses.
The coupling region of our device is designed as a coupling
capacitor: a part of the CPW resonator parallel and close to the
through line. Although $Q_{c}$ can not be calculated from coupling
strength analytically, numerical simulations show that a weak
coupling leads to a high quality factor. Therefore, the designed
$Q_{c}$ can be adjusted by changing either the length of the CPW
resonator parallelling to the through line or the distance between
them.
Certainly, the coupling strength should not be set extremely small,
otherwise the CPW resonator will not be excited for observation. Our
numerical results, considering our current fabrication techniques,
show that, if the geometry parameters for the coupling region are
set as those in Fig.~1(c), then the desired quality factor could be
up to $10^6$.

After the above design and relevant numerical simulations, we
fabricated our CPW resonator devices by magnetron sputtering and
photolithography technique.  Experimentally, a $160$ nm thick
niobium film was deposited on a $500$ $\mu$m thick silicon
substrate, and then contact exposure as well as reflect ion etching
are used to make the desired patterns on the film.

\begin{center}
\begin{overpic}[width=8cm]{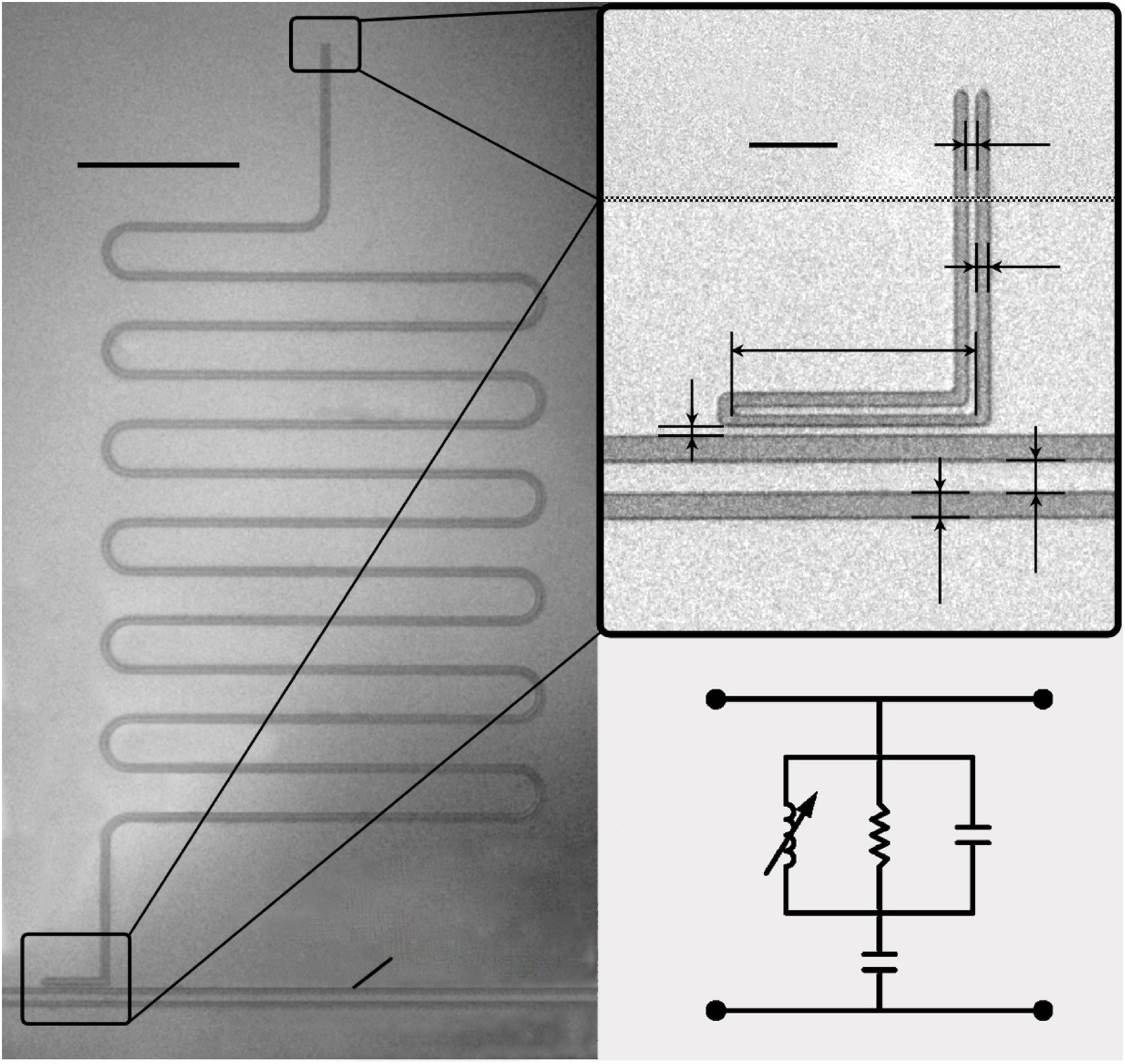}
\put(18,885){\bf (a)} \put(560,885){\bf (b)} \put(560,715){\bf (c)}
\put(560,322){\bf (d)} \put(800,420){$s_{t}$} \put(877,444){$w_{t}$}
\put(900,723){$s_{r}$} \put(895,830){$w_{r}$} \put(750,650){$l_{c}$}
\put(565,598){$g_{c}$}

\put(78,822){$\footnotesize 400\mu m$} \put(638,837){$\footnotesize
100\mu m$}
\put(270,105){Through line} \put(123,619){CPW Resonator}
\put(906,885){Nb} \put(648,180){L} \put(737,180){R} \put(815,180){C}
\put(703,71){C$_{c}$}
\end{overpic}
\end{center}
{\bf Figure 1. (a)} Microscope photograph of the key structure of
our device. Light and dark areas are niobium film and silicon
substrate, respectively. The through transmission line is used for
excitation and readout. The meandered CPW resonator has a overall
length of 15.814 mm, corresponding to a fundamental frequency around
1.8575 GHz. {\bf(b)} One end of the CPW resonator is shorted to the
ground plane. {\bf(c)} The magnified coupling region and main design
parameters: The length of the capacitor elbow $l_{c}$ is 160 $\mu$m
and the distance between through line gap and the capacitor elbow
gap $g_{c}$ is 7 $\mu$m. The width of center conductor $w_{r}$ and
the gap between it and the ground $s_{r}$ are both 7 $\mu$m.
{\bf(d)} The equivalent circuit of the measured device.

The device can be treated equivalently as a measured parallel lumped
RLC resonant circuit, shown in Fig.~1(d). where, $R, L,$ and $C$
characterize the small resistance, inductance, and capacitance of
the measured CPW resonator, respectively. $C_{c}$ is the coupling
capacitance between through line and the CPW resonator. Note that
the total inductance of the CPW resonator: $L=L_{m}+L_{k}$, consists
of two parts: the magnetic inductance $L_{m}$ and the kinetic
inductance $L_{k}$. In particular, the latter one depends strongly
on the temperature of the superconducting device. With such a
property, the CPW resonator could be used as an attractive candidate
of sensitive radiation detector.

The CPW resonator chip is glued (with GE varnish) onto a gold-plated
sample block made of oxygen-free high purity copper and wire bonded
to microwave transmission lines on a printed circuit board that was
carefully designed to ensure a good performance up to $6$ GHz
microwave signals. Additional bonds are placed at all chip sides for
good ground-connection.
During the measurements the sample block is placed at the mixing
chamber plate in our dilution refrigerator (whose base temperature
is around 20mK).
Fig.~2(b) shows the schematics of our measurement system. Cryogenic
coax cables connect the sample block to room temperature circuits.
Attenuators and DC blocks are positioned appropriately to reduce
unwanted circuit noises. The device is measured with the Agilent
E5071C vector network analyzer (VNA). The raw data collected
consists of complex transmission amplitude $S_{21}$ for $1601$
points in the desired frequency interval. Since the data is not
calibrated, losses in the coax cables are included.

\begin{center}
\begin{overpic}[width=5cm]{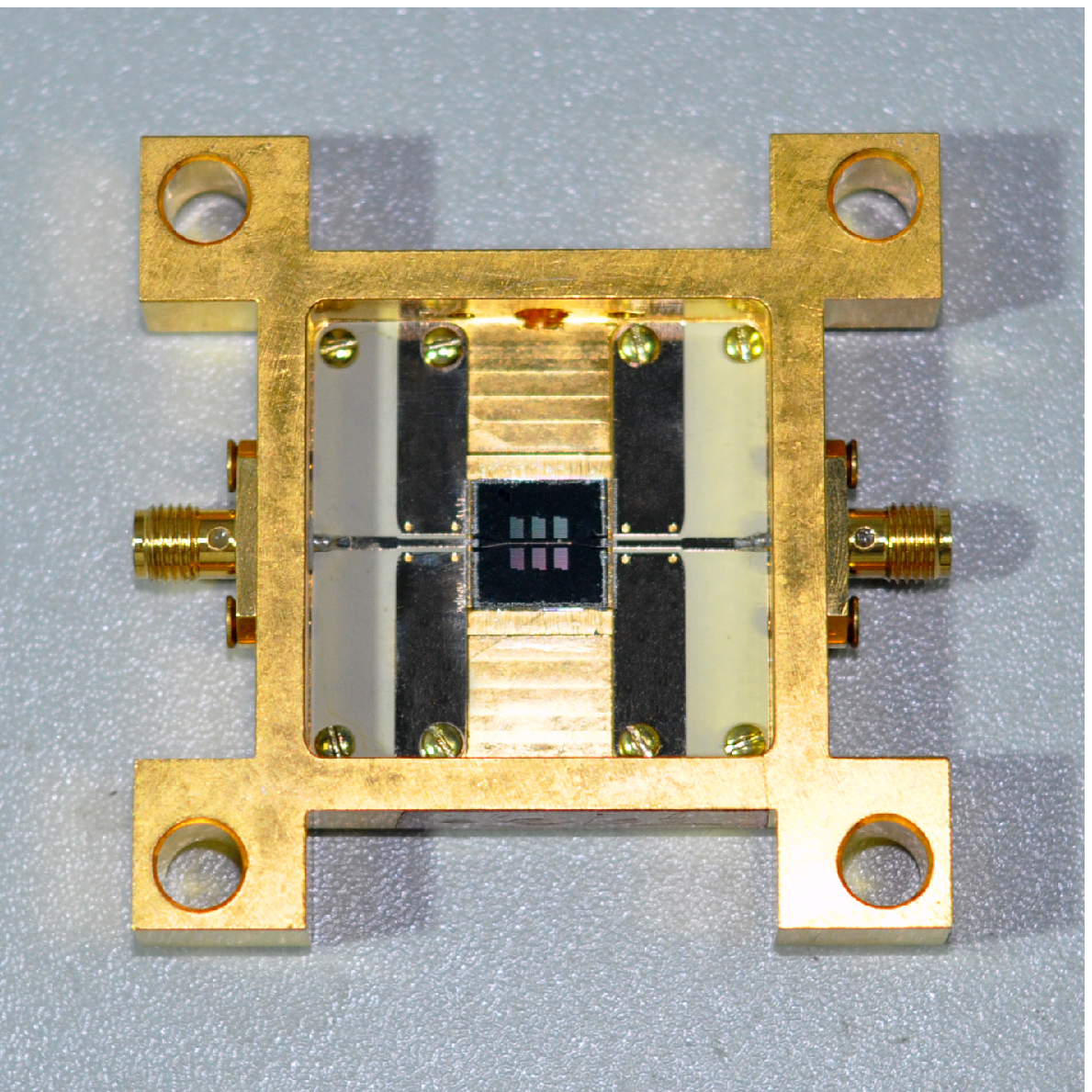}
\put(5,946){\bf (a)}
\end{overpic}
\end{center}
\begin{center}
\begin{overpic}[width=8cm]{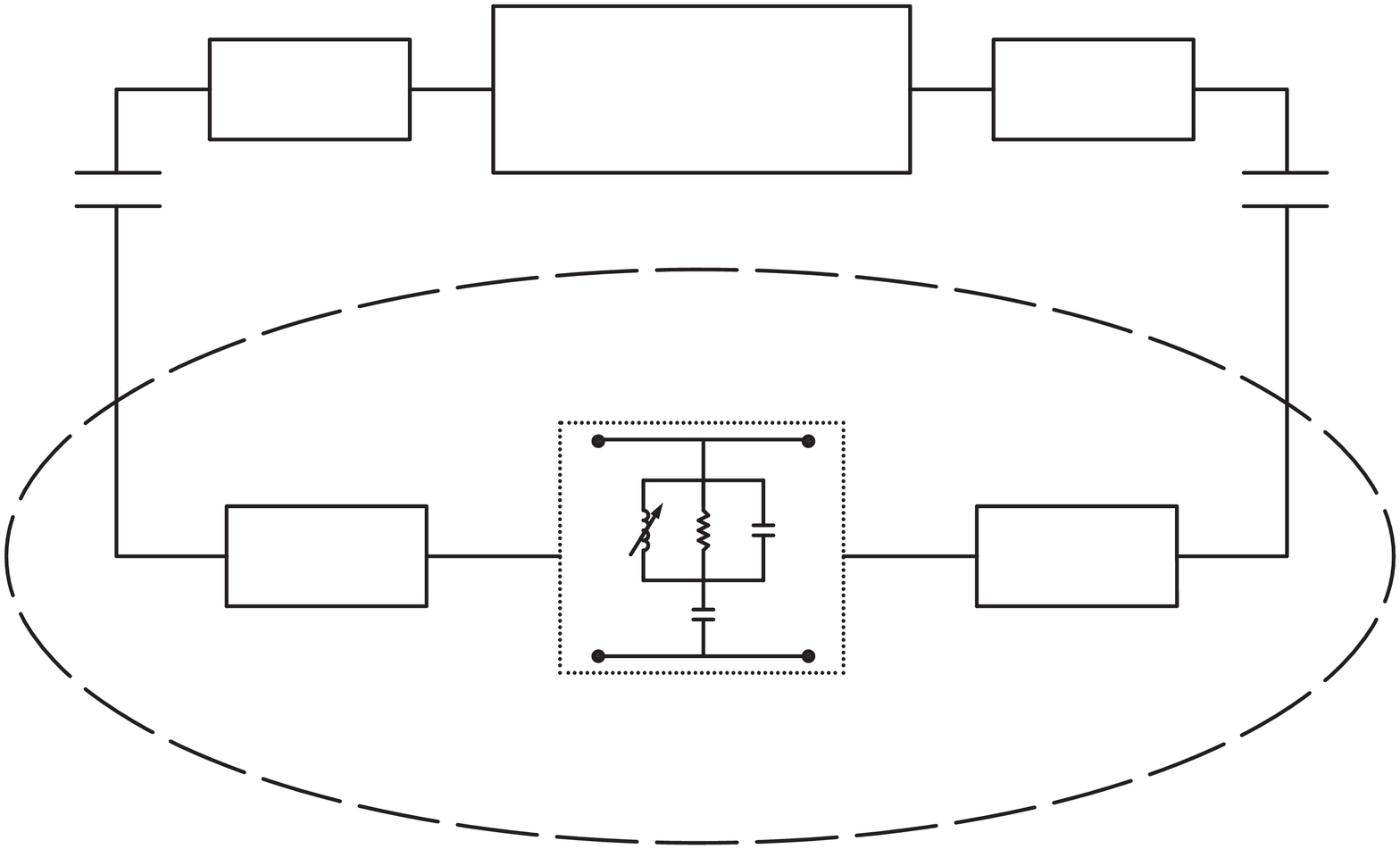}
\put(170,185){Attenuator} \put(660,185){Attenuator}
\put(455,145){DUT} \put(415,380){Cryogenic} \put(350,680){Room
temperature} \put(155,625){Attenuator} \put(673,625){Attenuator}
\put(455,550){VNA} \put(148,455){DC Block} \put(687,455){DC Block}
\put(30,680){\bf (b)}
\end{overpic}
\end{center}
{\bf Figure 2. (a)} Inner view of the sample block. {\bf(b)}
Schematics of the measurement system.\

\section{Experimental results}
By sweeping the frequency and adjusting the readout power of VNA,
the $S_{21}$ data were obtained and a resonance dip near the
designed position was identified. It is observed that the shape of
the resonance dip strongly depends on the driving power. For higher
power, various nonlinear effects reveal and the resonance features
of the device are complicated. While, within the lower power regime,
a maximum resonance dip depth was observed for an optimal driving
power~\cite{Power2}. Indeed, it is seen from Fig.~3 that, at the
driving power $-46$ dBm an optimal resonance dip with a sufficiently
high signal to noise ratio was found around 1.8575 GHz.
Note that the present data include all the losses in the coax
cables, which are estimated to be about -14 dB (at 1.8575 GHz) after
calibration. The inset of Fig.~3 shows a sharper dip, corresponding
to the same measurement with a wider bandwidth view.
\begin{center}
\begin{overpic}[width=8.5cm]{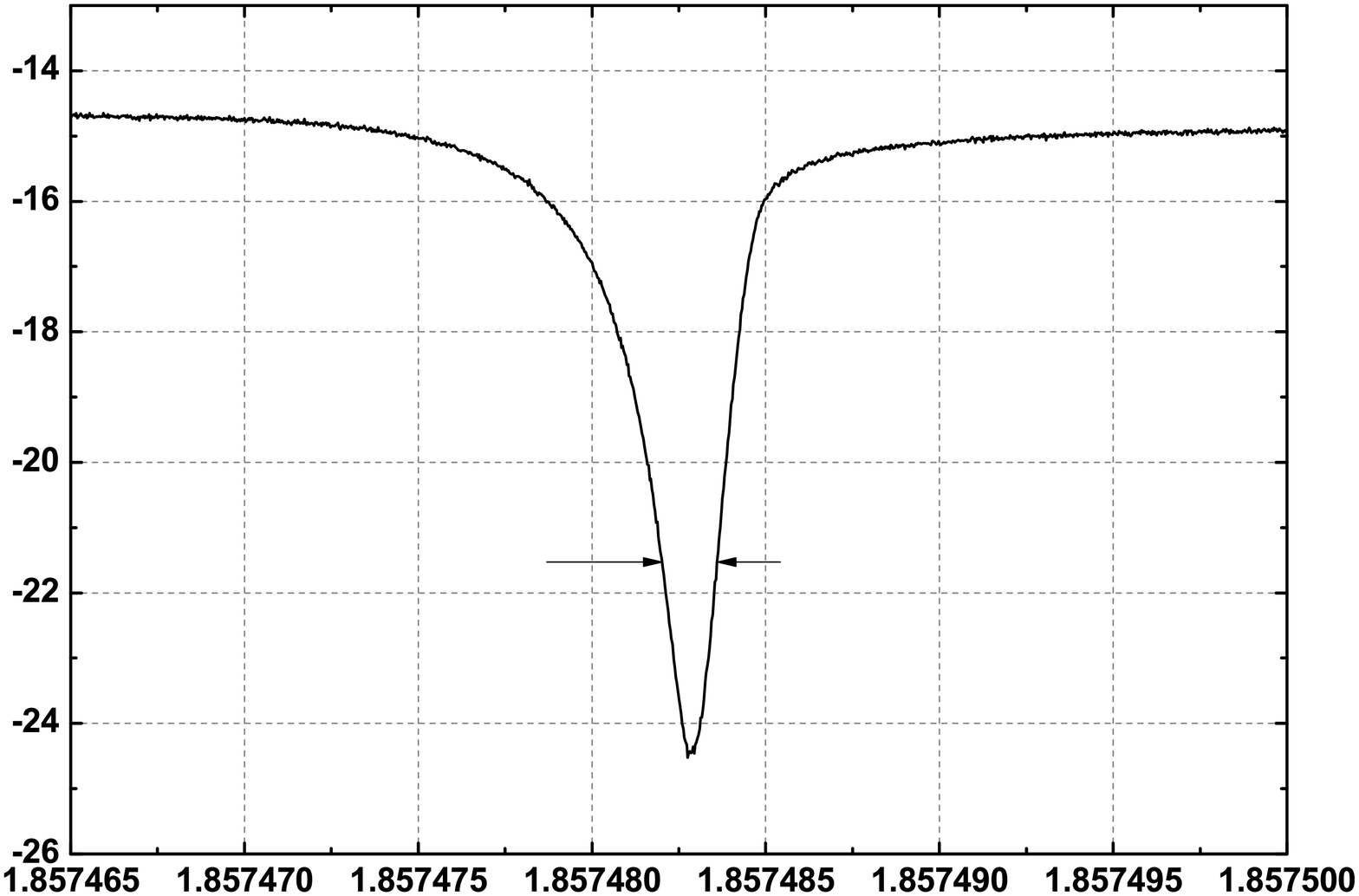}
\put(15,300){\rotatebox{90}{$\rm S_{21}$(dB)}}
\put(393,5){Frequency(GHz)} \put(180,320){\footnotesize T = 20 mK}
\put(456,300){\footnotesize $\Delta f$} \put(180,280){\footnotesize
$P_{VNA}$ = -46 dBm} \put(180,190){\footnotesize $\Delta f$ = 1.5937
KHz} \put(180,230){\footnotesize $f_{0}$ = 1.8574 GHz}
\put(180,140){\footnotesize $Q_{l} = 1.1654\times 10^{6}$}
\put(580,100){\includegraphics[width=3cm]{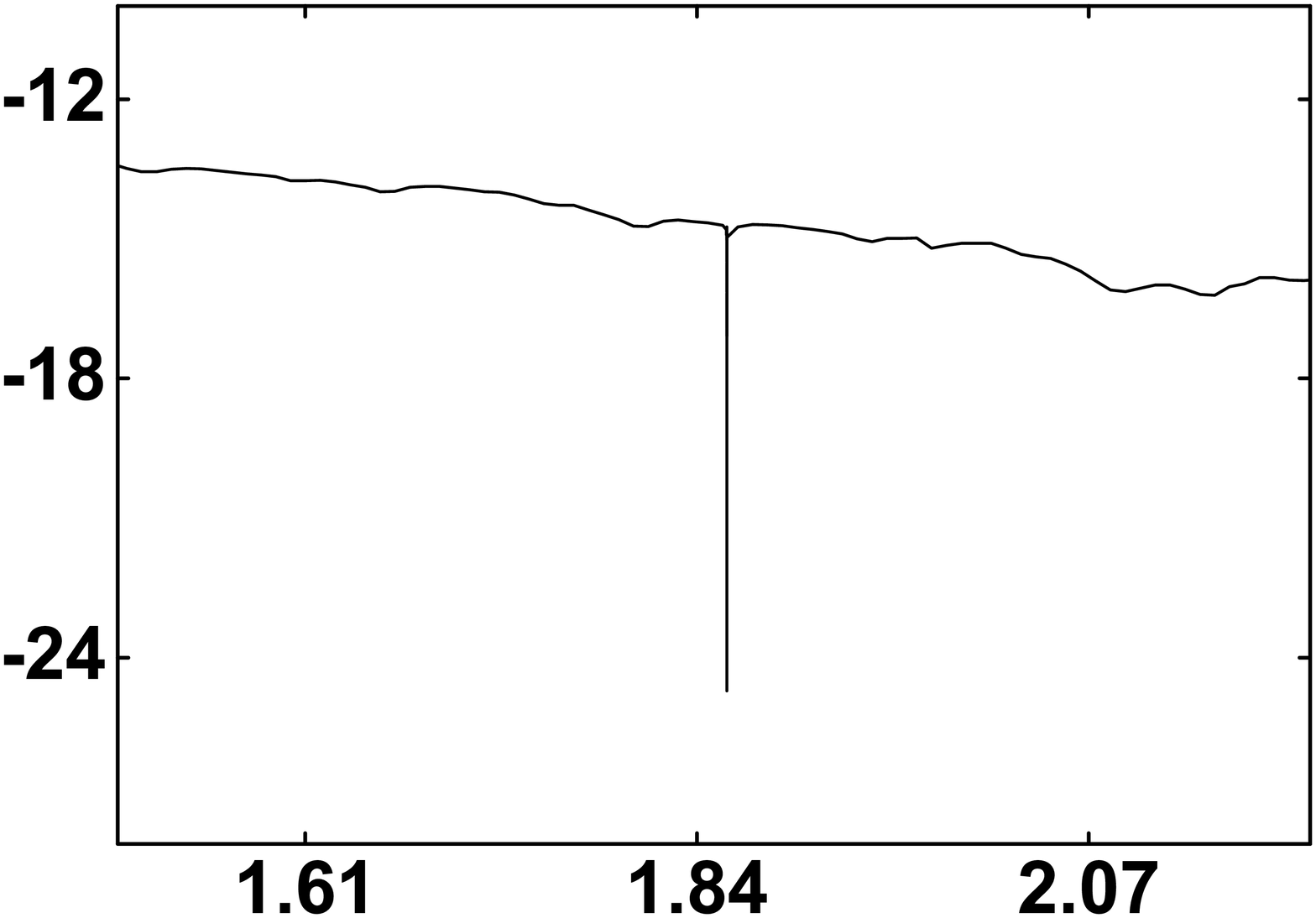}}
\end{overpic}
\end{center}
{\bf Figure 3.} The observed transmission dip. From the obtained
$S_{21}$ data, one can calculate the loaded quality factor of the
measured CPW resonator is $Q_{l}=f_{0}/\Delta f$ = 1.1654 $\times
10^{6}$. The inset shows a sharper resonance dip, corresponding to
the same measurements with a wider bandwidth view.

From the obtained $S_{21}$ data, the loaded (total) quality factor
of the CPW resonator can be calculated as
\begin{equation}
Q_{l}=\frac{f_{0}}{\Delta f},
\end{equation}
with $f_{0}$ being the resonance frequency and $\Delta f$ the
3dB-bandwidth. $Q_{l}$ includes both the source and load dissipation
of the device.
In our case, $f_{0}$ = 1.8575 GHz and $\Delta f$ = 1.5937 KHz were
observed. Thus, the loaded quality factor is calculated as $Q_{l} =
1.1654\times 10^{6}$.
Also, $Q_{i}$ and $Q_{c}$ can be individually calculated from the
minimal values of the measured $S_{21}$ parameters at the resonance
point, i.e., $Q_{i}=Q_{l}/|S'_{21}|$ and
$Q_{c}=Q_{l}/(1-|S'_{21}|)$, with $S'_{21}=min\{S_{21}\}$. For our
device, we found: $Q_{i}=3.2845\times 10^6$ and $Q_{c}=1.8063\times
10^6$.


Furthermore, we investigate experimentally the temperature
dependence of the resonance frequency and quality factor by slowly
enhancing the system temperature, e.g., from 20 mK to 1180 mK.
We can see clearly from Fig.~5 that, with increasing temperature,
the center frequency shifts to the higher value and the resonance
dip becomes broader and shallower.
Phenomenally, these behaviors can be explained in the framework of
the two-level systems (TLSs) theory~\cite{TLS1,TLS3}. In which the
unsaturated TLSs residing on the metal and dielectric surfaces are
believed to be a significant loss mechanism. The resonant
interaction of TLSs with electric field (microwave) leads to a
temperature dependent variation of the substrate dielectric
constant. As a consequence, both dielectric constant $\epsilon$ and
the surface inductance $L_{s}$ determine the resonance frequency as:
\begin{equation}
\frac{\delta f}{f}=\frac{1}{2} \frac{\delta
L_{s}}{L_{s}}-\frac{F}{2} \frac{\delta \epsilon}{\epsilon},
\end{equation}
where $F$ is a filling factor and $L_{s}=L_{m}+L_{k}(0)\times H(T)$.
The magnetic inductance $L_{m}$ remains essentially a constant in
the concerned temperature range. While, $H(T)$ represents the
temperature dependence of the kinetic inductance $L_{k}$, which is
strongly influenced by the temperature variation near the critical
temperature $T_{c}$.
However, at the present significantly-low temperature
$T<<T_{c}$(=9.2K for niobium), the kinetic inductance can still be
treated as a constant. Thus, the observed resonance frequency shifts
were mainly related to the temperature-dependent dielectric
constant. Indeed, it was shown that the dielectric constant
decreases with temperature, and thus the resonance frequency of the
CPW resonator increases.
Certainly, if the device works at the temperature closing to
$T_{c}$, then the effects due to TLSs saturate and the temperature
dependence of kinetic inductance is dominant. Therefore, the
increasing $L_{k}$ decreases the resonance frequency.
Basically, the surface resistance $R_{s}$ increases with temperature
monotonically, therefore the resonance dip is broader and shallower
(and thus the quality factor decreases) for the increasing
temperature.
\begin{center}
\begin{overpic}[width=8.5cm]{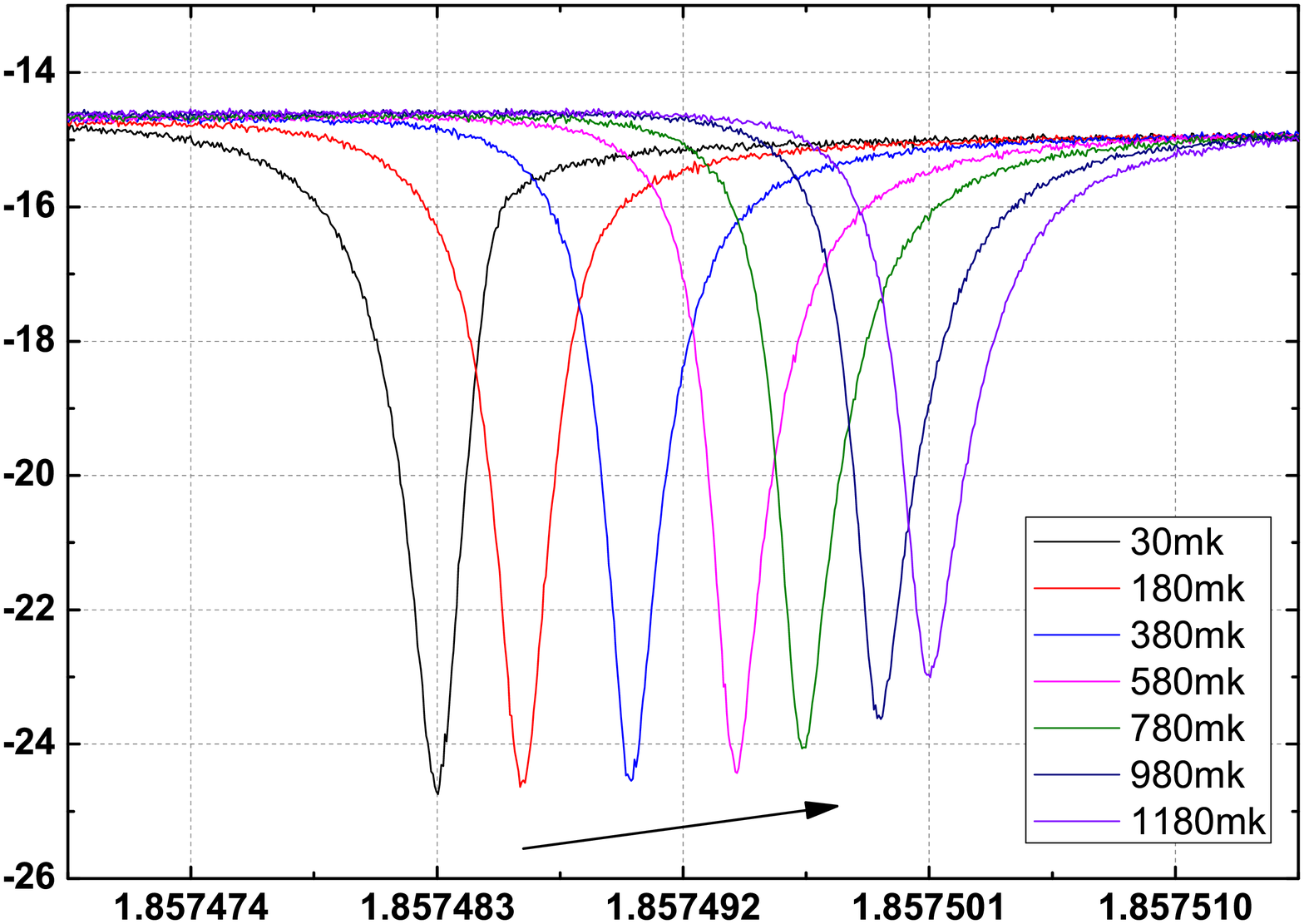}
\put(15,300){\rotatebox{90}{$\rm S_{21}$(dB)}}
\put(415,10){Frequency(GHz)}
\end{overpic}
\end{center}
{\bf Figure 4.} The increasing resonance frequency and decreasing
quality factor of the CPW resonator versus the increasing system
temperature for $T<<T_c.$

\section{Discussions and Conclusions}
The main advantage of superconducting CPW resonator device is that
it can be easily integrated. That is, multiple resonators operating
at different frequencies can be read out through a common feedline.
In fact, the number of the multiplexed resonators depends on several
factors, e.g., their quality factors and manufacture accuracy. In
principle, a device with higher-Q and better manufacture accuracy
allows for a smaller frequency separation and thus more resonators
multiplexed on one chip.

\begin{center}
\begin{overpic}[width=8.5cm]{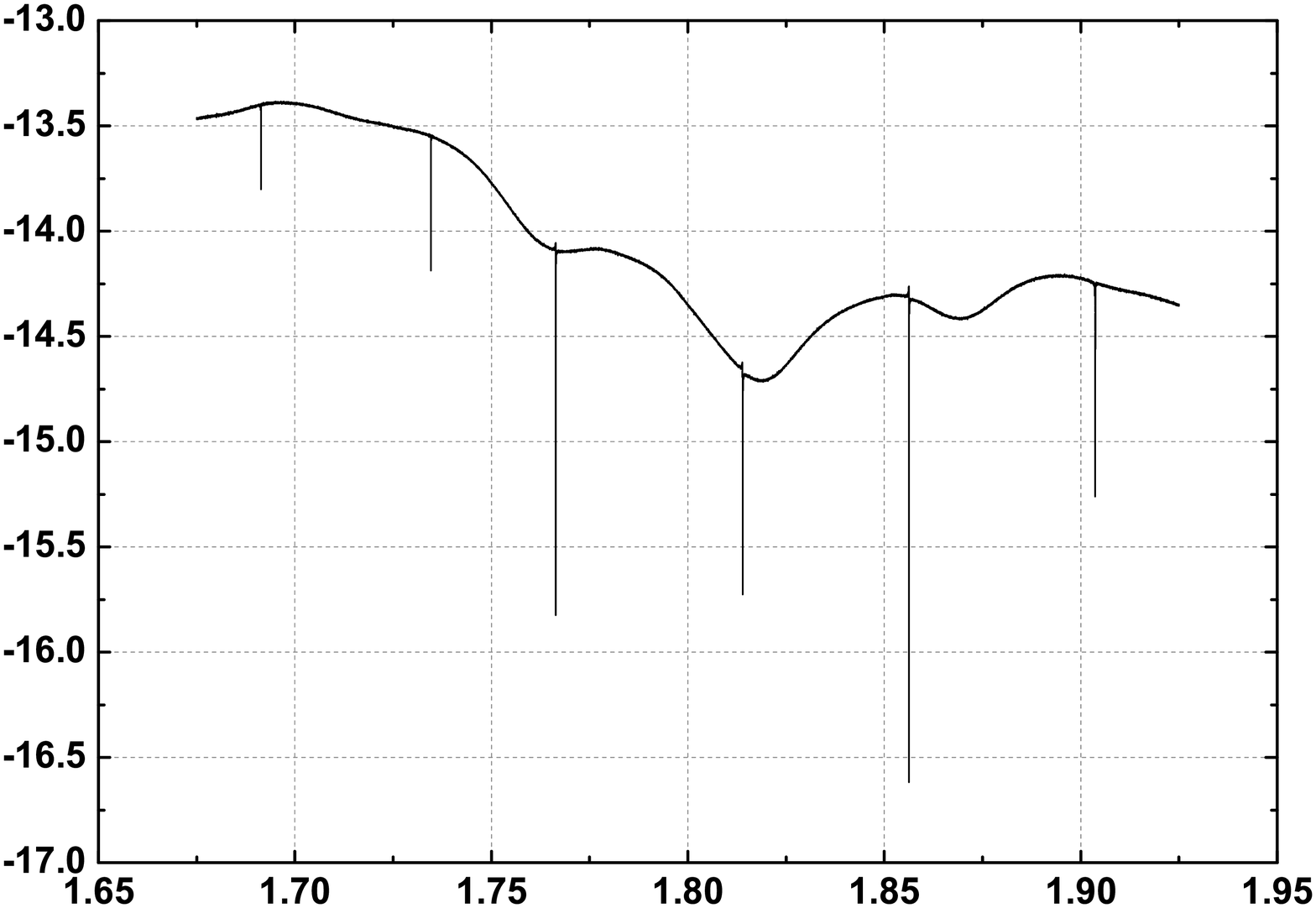}
\put(15,300){\rotatebox{90}{$\rm S_{21}$(dB)}}
\put(415,8){Frequency(GHz)}
\end{overpic}
\end{center}
{\bf Figure 5.} Resonance dips for six resonators multiplexed on a
chip.

In order to make a simple presentation of frequency domain
multiplexing scheme, we also fabricated a device with six resonators
coupled to a common through line. Each resonator is designed with a
different length, corresponding to a different resonance frequency.
Fig.~5 shows the measured results of such a multiplexed device,
where six resonance dips can be observed expectably.
One can see from the Table 1 that, the measured resonance frequency
$f_{m}$ are agrees well with the designed resonance frequency
$f_{t}$. Specifically, the designed length of the resonator No.5 is
the same as that measured in Fig.~3. The difference between the two
measured resonance frequencies is about $1.1$ MHz, which is very
small. This indicates that our manufacture is sufficiently accurate.

\begin{center}
\begin{tabular}{|c|c|c|c|}\hline
Number&$l_{d}(\mu $m$)$&$f_{t}($GHz$)$&$f_{m}($GHz$)$\\ \hline
1&17414&1.6958&1.6914 GHz\\ \hline 2&17014&1.7357&1.7346 GHz\\
\hline 3&16614&1.7775&1.7664 GHz\\ \hline 4&16214&1.8213&1.8139
GHz\\ \hline 5&15814&1.8665&1.8563 GHz\\ \hline
6&15414&1.9159&1.9037 GHz\\ \hline
\end{tabular}
\end{center}
{\bf Table 1.} Designed and measured resonance frequencies of six
CPW resonators on a chip.
\\
\\
In summary, we have designed and fabricated a high-quality niobium
quarter-wave CPW resonator. The transmission of microwave is
measured with a vector network analyzer and a sharp resonance dip is
observed at designed position. At 20 mK, the measured quality factor
of the device is $Q_{l}$ = 1.1654 $\times 10^{6}$, which is
significantly high and could be applied for sensitive radiation
detections and quantum information processing.
\\
\\
{\bf Acknowledgments.} This work was supported in part by the
National Science Foundation grant Nos. 90921010, 11174373, and the
National Fundamental Research Program of China through Grant No.
2010CB923104.

\newpage

\end{document}